\documentstyle[aps]{revtex}
\begin{document}

\title{Functional statistical inference of parton distributions}
\author{Vipul Periwal}
\address{Department of Physics,
Princeton University,
Princeton, New Jersey 08544}

\def\dd{\hbox{d}}
\def\tr{\hbox{tr}}\def\Tr{\hbox{Tr}}
\def\ee#1{{\rm e}^{{#1}}}
\def\DD{\hbox{D}}
\def\pdf{parton distribution function}
\def\part{\partial}
\def\la{\lambda}
\maketitle
\begin{abstract} Bialek, Callan and Strong have recently given a
solution of the problem of determining a continuous probability
distribution from a finite set of experimental measurements by
formulating it as a one-dimensional quantum field theory.  This report
applies an extension of their
formalism to the inference of functional parton distributions from
scattering data.
\end{abstract}
The parton model is applicable to large momentum transfer hadronic
processes.  Scaling in hadronic cross-sections is interpreted in the
parton model as a consequence of the existence of charged pointlike
constituents in hadrons.  Parton model cross-sections are
calculated by combining amplitudes for the scattering of these
constituents with probability densities (\pdf s)
for finding a given constituent in a hadron
carrying  a given fraction of the total momentum of the hadron.

In perturbative QCD\cite{review}, scattering cross-sections can be given a
parton model--inspired form, by invoking a factorization scale,
$\mu_{f}.$
Roughly, it is supposed that all internal lines with momenta off-shell by
more than $\mu_{f}$ are included in the amplitude factor of the
cross-section, and the effects of all lines with softer momenta are
included in
the parton distributions.  (The precise meaning of
$\mu_{f}$ is incorporated in the operational definition  of 
\pdf s via factorization schemes, and will not be needed here.)
The latter quantities cannot, at present,
be derived from QCD, and must therefore be experimentally determined.
There is an important  theoretical constraint on these distributions:
No physical quantity can depend on the choice of factorization scale,
thus the calculable $\mu_{f}$ dependence of the amplitude factors
can be translated into the $\mu_{f}$ dependence of the distribution
functions, leading to the
Gribov--Lipatov--Altarelli--Parisi (GLAP)
evolution equations satisfied by the \pdf s.
Knowledge of the \pdf s at some value
of $\mu_{f}$ therefore is input for the prediction of hadron scattering
cross-sections at other energies.  Thus
the importance of inferring these \pdf s cannot be
over-emphasized---the prospect of extracting new physics from hadron
collider experiments requires the subtraction of known hadronic
physics to a high degree of sprecision.

Besides systematic experimental uncertainties, the determination of
\pdf s is a difficult task because the fitting procedure is
of necessity somewhat circuitous\cite{review}.  Given the \pdf s at some
momentum transfer, they must be evolved to other values using
the GLAP equations.  The experimental data is then fit to
cross-sections computed from the set of evolved \pdf s. Then
the initial \pdf s are altered to improve the experimental fit, and
the whole set of steps is repeated.  Furthermore,  the functional form of
the parton distributions is unknown---we do not know how to solve
QCD.  Thus a fit to parameters in a postulated functional form
is beset with the worry that the true functional form of the distribution is
not the postulated one, in which case it is quite probable that the
true distribution does not even lie on the manifold of parameters at all.
That this is not a far-fetched scenario is pointed out by the
ln~$x$ dependence that appears to smooth  the $\mu_{f}$ dependence of
$a,b$ in the  
original $x^{a}(1-x)^{b}$ form\cite{review}.

The aim of this report is to show that a recent formulation of
the problem of functional statistical inference due to Bialek, Callan
and Strong\cite{bialek}
can be simply extended to provide a framework for parametrization
independent inference of parton distribution functions.

My attention was drawn to this problem by
G.~Sterman who   emphasized the importance of a
parametrization-independent fit.  Further, after this work was
completed, and presented at several seminars, M.~Peskin drew my
attention to the work of Giele and Keller\cite{giele}.  These authors
point out the utility  of Bayesian methods in the inference
of parton distributions\cite{giele}, motivated by the problem of
assigning error estimates to inferred \pdf s.   It seems
therefore that the present report contains material that  
complements \cite{giele}, and  fills a gap mentioned in
the concluding section of \cite{giele}.

I review briefly  the  elegant formulation of
the problem of inferring a continuous probability distribution from a
finite number of data points
given  by Bialek, Callan and Strong\cite{bialek}.  A small
variation on their work, to take into account the fact that
probability distributions are densities, was given in \cite{me,metwo}, but
I shall not bother with such niceties here.  I then extend the work
of \cite{bialek} by constructing an exact 
solution of the equations obtained in \cite{bialek}\ for a finite
data set, instead of the WKB type analysis given in \cite{bialek}.  
With this material in hand, the \pdf\ problem involves just a few
conceptual extensions.

Ref.~\cite{bialek}\ used Bayes' rule to write the
probability of the probability distribution $Q,$ given the data
$\{x_{i}\},$ as
\begin{eqnarray}
P[Q(x)| x_1 , x_2 , ... , x_N]
\,
 =
{{P[x_1 , x_2 , ... , x_N | Q(x)] P[Q(x) ]}
\over {P(x_1 , x_2 , ... , x_N )}} \, =
{{Q(x_1 ) Q(x_2 ) \cdots Q(x_N) P[Q(x)]}
\over{\int [dQ(x)] Q(x_1 ) Q(x_2 ) \cdots Q(x_N) P[Q( x)]}} ,
\label{conditional}
\end{eqnarray}
where the factors $Q(x_{i})$ arise because each $x_{i}$ is chosen
independently from the distribution $Q(x),$ and $P[Q]$ encodes the
{\it a priori} hypotheses about the form of $Q.$     The optimal
least-square estimate of $Q,$ $Q_{{\rm est}}(x,\{x_{i}\}),$ is then
\begin{equation}
Q_{\rm est} (x; \{x_i\} )
=
{{\langle Q(x) Q(x_1 ) Q(x_2 ) \cdots Q(x_N) \rangle ^{(0)}}\over
{\langle Q(x_1 ) Q(x_2 ) \cdots Q(x_N) \rangle^{(0)}}} ,
\label{est}
\end{equation}
where $\langle \cdots \rangle^{(0)}$ denotes an  expectation value with
respect to the {\it a priori} distribution $P[Q(x)]$.

In this field-theoretic setting,  Bialek et al.~\cite{bialek} assumed that
the prior distribution $P[Q]$ should penalize large gradients, so
written in terms of an unconstrained variable $\phi \equiv -\ln (\ell
Q) \in (-\infty,+\infty),$ they assumed
\begin{eqnarray}
P_\ell[\phi(x)]
 =  {1\over Z} \exp\left[ -{\ell\over2} \int dx (\partial_x \phi)^2
\right]
\, \times
\delta \left[ 1 - {1\over \ell} \int dx \ee{-\phi(x)}\right] ,
\end{eqnarray}
with $\ell$ a parameter that they later averaged over independently.
This form for the prior distribution is very simple, and quite
minimal in terms of underlying assumptions, so
conclusions drawn from it should be robust.

Write $Q(x) \equiv   \exp(-\phi(x))/\ell,$
with
\begin{eqnarray}
P[\phi] = {1\over Z} \exp\left[ -{\ell\over2} \int dx
 (\partial_x \phi)^2
\right] \delta \left[ 1 - {1\over \ell} \int dx  \ee{-\phi(x)}\right] ,
\end{eqnarray}
Now, we want to evaluate
\begin{equation}
 \langle Q(x_1 ) Q(x_2 ) \cdots Q(x_N) \rangle^{(0)} 
=
\int \DD\phi
P[\phi] \prod_{i=1}^N \exp[-\phi(x_i)] \\
  =
 {1\over Z} \int {{\dd\lambda }\over{2\pi}}
\int \DD\phi
\exp\left[ - S(\phi; \lambda ) \right] ,
\label{problem}
\end{equation}
where
\begin{eqnarray}
S(\phi; \lambda ) =
{1\over2} \int \dd x (\partial_x \phi)^2
\,
+i{\lambda } \int \dd x  \ee{-\phi(x)}
+\sum_{i=1}^N \phi (x_i ) - i\lambda.
\label{accion}
\end{eqnarray}
Ref.~\cite{bialek} showed that the integral in eq.~\ref{problem}\
can be evaluated by saddle-points  for $N$ large, and showed that the
best estimate for the inferred distribution tends to the true distribution.
The width on the space of distributions (in other words, error bars
about the distribution) can also be computed in this formulation.

Since I will use a slightly different analysis for
the \pdf\ problem, adapted from \cite{metwo} (see 
eq.'s~\ref{solute},\ref{continuity},\ref{jump},\ref{eval}\ below),
I first recast  their analysis\cite{bialek} in the 
framework I need.  Finding the configuration which extremizes 
the classical action, eq.~\ref{accion},  
is equivalent to maximum likelihood estimation,  which
chooses the distribution, $Q(x)$, that maximizes $P[Q(x) |\{x_i\}].$
The classical equations of motion for $\phi$ and $\lambda$ are
\begin{eqnarray}
\ell \partial_x^2 \phi_{\rm cl}(x) +i{\lambda_{\rm cl}\over\ell}
\ee{-\phi_{\rm cl}(x)} &=& \sum_{i=1}^N \delta (x-x_i)
\label{phieqn}
\\
{1\over\ell} \int \dd x \ee{-\phi_{\rm cl}(x)} &=& 1 .
\label{lambdaeqn}
\end{eqnarray}
Eq.~\ref{phieqn}\ and eq.~\ref{lambdaeqn}\
imply $i\lambda_{\rm cl} = N$,
provided that $\partial\phi (x)$
vanishes as $|x| \rightarrow \infty.$ 

It is easy to solve eq.~\ref{phieqn}.  
Notice that the equation
\begin{equation}
 	\partial_x^2 f(x) +N \ee{-f(x)} =0
\end{equation}
is solved by
\begin{equation}
f(x;a,b) = 2\ln\left({\sinh(a(x-b))\over a}\right) -\ln {2\over  N}.
\label{solute}
\end{equation}
Assume first that
the data points are all distinct, and let $\ell=1$ for the time 
being---it is easily restored by dimensional analysis. 
We order the data points $\{x_{i}\}$ in ascending order $x_{i}<x_{i+1}.$
On the interval $(x_{i-1} ,x_{i}],$ with $x_{0}\equiv 
-\infty, x_{N+1}\equiv 
+\infty,$  we take the solution $\phi_{{\rm cl}} = 
f(x;a_{i-1},b_{i-1}) .$  $a_{0}=0=a_{N} $ is implied by the 
asymptotics above.  Remark that $f=
\ln (x-b)^{2} - \ln {2\over  N} $ is the limiting case when
$a\downarrow 0.$  At $x_{i}, i=1,\ldots,N$ we impose the matching conditions
\begin{equation}
{\sinh(a_{i-1}(x_{i}-b_{i-1}))\over a_{i-1}} = 
{\sinh(a_{i}(x_{i}-b_{i}))\over a_{i}}
\label{continuity}
\end{equation}
which makes $\phi_{{\rm cl}}$ continuous, and 
\begin{equation}
2{\sinh(a_{i-1}(x_{i}-b_{i-1}))\over a_{i-1}} = 
{\cosh(a_{i}(x_{i}-b_{i})) }	- {\cosh(a_{i-1}(x_{i}-b_{i-1})) }
\label{jump}
\end{equation}
which accounts for the delta functions at $x_{i}.$   
There are $2N$ matching conditions, and there are  $2N$ free
parameters, since there are $N+1$ intervals, but $a_{0}$ and $a_{N}$
have been fixed by the asymptotics already.  If data points 
co\"\i{}ncide, the number of intervals is appropriately reduced, and 
the right hand side of \ref{jump}\ is multiplied by the number of 
data points at that value of $x_{i}.$  Thus we have an explicit
exact solution of the variational equations valid for arbitrary $N.$
The \pdf\ problem will require assuming that the probability 
distribution has finite support on the unit interval, but the changes 
required in these formul\ae\ for this case are rather minimal, and are
explained in \cite{me,metwo}.

What we really want to do, of course, is to compute eq.~\ref{problem}.
As in \cite{bialek}, but with the exact solution 
$\phi_{\rm cl}(x;\{x_{i}\})$ constructed 
above, the one-loop contribution is  
\begin{equation}
S_{1}={1\over 2}\left({N\over{\ell}}\right)^{1/2}
\int \dd x   e^{-{1\over 2}\phi_{\rm cl}(x)} ,
\end{equation}
up to terms that are lower order in powers of $N.$  (Some care is 
needed in taking into account the fluctuations in $\lambda,$ since 
these mix with constant shifts in $\phi,$ but this is not difficult.) 
So finally, we note that 
\begin{equation}
 \langle Q(x_1 ) Q(x_2 ) \cdots Q(x_N )\rangle^{(0)}
\approx \prod_{i=1}^N \ee{-\phi_{{\rm cl}}(x_{i})}\times
\exp [-{\ell\over2}\int (\part_{x}\phi_{{\rm cl}})^{2} - S_{1}] ,
\label{eval}
\end{equation}
up to higher loop corrections.  While $\part_{x}\phi_{{\rm 
cl}}$ is not continuous, it  is square-integrable, so this
expression is well-defined.   The last step in the problem is to 
determine $\ell.$  This is explained in \cite{bialek}\ from a `global'
perspective, with $\ell$ constant over the whole support of the 
distribution, and a variant is explained in \cite{me,metwo}\ from
a local perspective.

Turning now to the \pdf\ problem,
there are two new complications beyond the evaluation of
eq.~\ref{problem}.   As mentioned above, \pdf s
are inputs into the determination of cross-sections.  These
cross-sections, in turn, are the quantities that are experimentally
measured.  Thus the problem of inferring \pdf s is a Bayesian problem
with a different prior distribution compared to the simple direct
inference problem addressed in \cite{bialek}.  The prior distribution
favors continuous \pdf s, but the actual probabilities (in other words,
the cross-sections) are (in general) nonlinear  functions of the \pdf{s}.
For example, for the Drell-Yan process
\begin{equation}
\sigma_{{AB}}(p,p')	= \sum_{i,j}\int_{0}^{1}\int_{0}^{1}\dd x\dd x'
\hat\sigma_{ij}(xp,xp',\mu;\mu_{f}) \phi_{{i|A}}(x,\mu;\mu_{f})
\phi_{j|B}(x',\mu;\mu_{f})
\end{equation}
is the cross-section, with $\phi_{j|B}$ the probability of
finding a parton of type $j$ carrying fraction $x$ of the
total momentum in a hadron $B,$ and
$\hat\sigma_{ij}(xp,xp',\mu;\mu_{f})$ is the cross-section for
scattering partons of type $i,j.$  Here the dependence on the
factorization scale and the normalization scale have been indicated
explicitly.  While the dependence of the cross-section on the \pdf{s}  is
nonlinear in this example, there are also examples where the
dependence is linear, for example, fragmentation cross-sections
\begin{equation}
	\dd\sigma_{C}(l)=\sum_{i}\int_{0}^{1} \dd x \
\dd\hat\sigma_{i}(l/x,\mu;\mu_{f}) D_{{C|i}}(x,\mu;\mu_{f}) ,
\end{equation}
where $D_{C|i}$ is the fragmentation function for parton $i$ to
produce a hadron $C$ with momentum $l.$
Furthermore, the \pdf s satisfy the GLAP equations, which are linear
equations.  Thus we do not integrate over \pdf s at all values of
$\mu_{f},$ but rather at a chosen value, say $\mu_{0},$ with
\pdf s at other values determined by the GLAP equations
\begin{equation}
	\mu^{2}{\part \over {\part \mu^{2}}} \phi_{{i|A}}(x,\mu;\mu )
	=\sum_{j}\int_{x}^{1} {\dd\xi\over \xi}
	P_{ij}[x/\xi,\alpha_{s}(\mu^{2})] \phi_{{j|A}}(\xi,\mu;\mu_{f}) .
	\label{glap}
\end{equation}
These equations are   a consequence of factorization and the fact that
physics is independent of the choice of factorization scale.

This is an  outline of the situation we wish to study.  Let us
abstract from these facts the problem we need to solve.  We want to
determine the values of some continuous fields
$\phi^{i}(x,t),$  given that these fields determine some families of
probability
distributions $Q_{\alpha}[\phi](t).$  We are given a finite set of
experimental data
$\{Q_{\alpha_{I}}(t_{I})\},$   with the $\alpha_{I}$ the type of
cross-section observed for data point $I,$ and $t_{I}$ the momentum
transfer for this data point.  We are normalizing 
cross-sections as appropriate for probability distributions.
If the GLAP equations were unknown, we
would have to determine $\phi^{i}(x,t)$ for all values of $t,$ and the
problem would involve a two-dimensional field theory as opposed to
the quantum mechanics problem, eq.~\ref{problem}.
As it
happens, the GLAP equations are exact to all orders in perturbation
theory\cite{review}.  We need only determine $\phi^{i}(x,t_{0})$
at a given value of $t_{0}.$  We can write the solution of 
eq.~\ref{glap}\ as 
\begin{equation}
	\phi_{i}(x,t) = \int_{0}^{1} \dd x' K_{ij}(x,x';t,t_{0}) \phi_{j}(x',t_{0}).
\end{equation}

In this abstract formulation, the $\phi_{i}$ need not be positive or
normalized in
any sense.  For the actual physical problem, the $\phi_{i}$
themselves are probability distributions.  This is the point where the
{\it choice} of a Bayesian prior affects the width of the distribution on
the space of distributions, in other words, affects the error bars
around the inferred \pdf{s}.  The choice of prior also affects the
convergence to the `true' \pdf{s} as the sample size becomes larger.
To be concrete, we observe that eq.~\ref{glap} is naturally expressed
in terms of $\phi,$ whereas  for calculation in
the functional integral  we are formulating,  the natural variable to
use is $\ln\phi,$ which is not constrained to be positive definite.
Assigning a priori probabilities that favour continuity of $\phi$
leads to a measure in the functional integral weighted by
$\int (\part_{x}\phi)^{2},$ whereas favoring the continuity of
$\ln\phi$ leads to $\int \phi_{i}^{-2}(\part_{x}\phi_{i})^{2}.$
There seems to be no way to decide on the appropriate Bayesian prior
without actual data analysis with different priors---this caveat
applies to practically all Bayesian statistical inference.

To be explicit, I consider each prior in turn.  For either  choice,
our problem is to evaluate
\begin{equation}
	\int  \dd\mu \prod_{i}\delta\left(\int_{0}^{1} 
	\dd x\phi_{i}(x,t_{0}) - 
	1\right)
	\prod_{I}Q_{\alpha_{I}}(t_{I}) \times \exp(-S_{0}),
	\label{pdfproblem}
\end{equation}
with $S_{0}= {1\over 2}\int\dd x\sum_{j}  
	(\part_{x}\phi_{j}(x,t_{0}))^{2}$ with the measure
	$\dd\mu\equiv \DD\phi_{i}(t_{0}),$ and the restriction $\phi_{i}\ge0,$
	or $S_{0}= {1\over 2}\int\dd x\sum_{j}  
	(\phi_{j}^{-1}\part_{x}\phi_{j}(x,t_{0}))^{2} $ with the measure
	$\dd\mu\equiv \DD\ln\phi_{i}(t_{0}).$
Eq.~\ref{pdfproblem}\ 
is obviously of the same variety as   eq.~\ref{problem}.  There 
is no conceptual or computational problem in incorporating other 
priors if the physics warrants such changes.  
The variational equation determining the maximum likelihood 
distributions is 
\begin{equation}
	\part_{x}^{2}\phi_{j}(x,t_{0}) - i\la_{j} = - \sum_{I,k}\int_{0}^{1}\dd 
	x' {{\delta 
	\ln Q_{\alpha_{I}}(t_{I})}\over \delta\phi_{k}(x',t_{I})} 
	K_{kj}(x',x;t_{I},t_{0}) 
	\label{former}
\end{equation}
for the former choice of $S_{0},$ and 
\begin{equation}
\part_{x}\phi_{j}^{-2}\part_{x} 
\phi_{j}(x,t_{0}) -i\la_{j} = - \sum_{I,k}\int_{0}^{1}\dd 
	x' {{\delta 
	\ln Q_{\alpha_{I}}(t_{I})}\over \delta\phi_{k}(x',t_{I})} 
	K_{kj}(x',x;t_{I},t_{0}) 
	\label{latter}
\end{equation}
for the latter choice.  Here $\la_{j}$ are Lagrange multipliers, 
which ensure that $\int\phi_{j}\dd x = 1.$
It is simpler to write eq.~\ref{latter}\ in 
terms of $\psi_{j}\equiv-\ln\phi_{j}:$
\begin{equation}
\part_{x}^{2}
\psi_{j}(x,t_{0}) +i\la_{j}\exp(-\psi_{j}) = \sum_{I,k}\int_{0}^{1}\dd 
	x' {{\delta 
	\ln Q_{\alpha_{I}}(t_{I})}\over \delta\psi_{k}(x',t_{I})} 
	K_{kj}(x',x;t_{I},t_{0}) 
	\label{latterprime} 
\end{equation}
since we do not have to constrain $\psi_{k}$ to be positive, unlike 
$\phi_{k}.$  
These are the general forms appropriate for obtaining \pdf s from  a 
global fit to all available data, including different physical 
processes in one fell swoop.  Obviously one can restrict the data
points included to be {\it e.g.} deep inelastic scattering events 
alone, or Drell-Yan events alone and compare \pdf s so 
obtained\cite{review}. 

Of course, this global fit may not be desirable in all instances since 
it is (usually) inappropriate to combine data-sets with different systematic
errors, but it may be helpful to have a formula available that 
explicitly implements  a global fit, perhaps as a check.
Further, I cannot give an analytical solution to eq.~\ref{latterprime}\
analogous to eq.~\ref{solute}.  This is not terribly disturbing since
numerical methods 
are required for solving the matching equations  
(eq.'s~\ref{continuity},\ref{jump}) in any event.
I do not discuss 
error estimations in depth, 
since they were the focus of the analysis in \cite{giele}.
Suffice it to say that in Bayesian approaches,  what one
calculates is the probability that a given set of \pdf s is the true 
set.  The width of the distribution on the space of \pdf s
can be extracted without much effort by computing the effective action
corresponding to the functional integral in eq.~\ref{pdfproblem}, not 
just the effective action at a stationary point.

An interesting possibility that one could also address in this 
framework is the following:  The GLAP equations are valid to all orders
in perturbation theory.  One might wonder if there is any experimental 
evidence for non-perturbative 
deviations from the GLAP equations.  This is easily 
incorporated in the present framework in several possible ways.  
Basically, instead of imposing the GLAP equations by hand, we need to
integrate over all $\phi_{i}$ independently at all values of $t,$ and
impose an additional term in the action ${1\over 2}\int\dd t\dd x\sum_{j}  
	(\phi_{j}^{-1}\part_{x}\phi_{j}(x,t))^{2}  $ of the general form
\begin{equation}
	\int\dd x\dd t\sum_{i} \left({\part\over\part{t}}\phi_{i}(x,t) -
	\int_{x}^{1} {{\dd x'}\over x'} \sum_{j}
	{\rm GLAP}_{{ij}}(x/x';t ) \phi_{j}(x',t)\right)^{2}
\end{equation}
where  ${\rm GLAP}_{{ij}}(x/x';t ) $ stands for the operator that 
appears on the right hand side of eq.~\ref{glap}.  Such a term 
implements a competition between smoothness and satisfying 
eq.~\ref{glap}\ in the Bayesian sense.  

There are various aspects of the suggestion in this report that need 
to be carefully considered before the suggested  framework will be of any 
value.  The problem of determining \pdf s is not a theorist's 
concoction, but a practical one.  As such, a great advantage is that
the proposed
scheme can be tested with real data, and rejected or accepted accordingly.

I am grateful to George Sterman for his suggestion,
to Curt Callan for encouraging comments, and to Michael Peskin for
drawing my attention to \cite{giele}.
This work was supported in part by NSF grant PHY96-00258.

\end{document}